\documentclass[12pt,preprint]{emulateapj}

\usepackage{color}

\newcommand{\feoh}{\textrm{[Fe/H]}}

\newcommand{\triplea}{triple-$\alpha$ reaction}
\newcommand{\nucm}[2]{\ensuremath{{}^{#1}{\rm #2}}}
\newcommand{\msun}{\ensuremath{M_{\odot}}}
\newcommand{\lsun}{\ensuremath{L_{\odot}}}

\newcommand{\pow}[2]{\ensuremath{{#1} \times 10^{#2}}}
\newcommand{\hyp}{\ensuremath{\, \mathchar`- \,}}
\newcommand{\sit}{{\it s}}

\slugcomment{accepted 2011 July 25}

\shorttitle{Stellar Evolution Constraints on the Triple-$\alpha$ Reaction Rate}
\shortauthors{Suda et al.}

\begin{document}


\title{Stellar Evolution Constraints on the Triple-$\alpha$ Reaction Rate}


\author{Takuma Suda\altaffilmark{1,2,3}, Raphael Hirschi\altaffilmark{2,4}, and Masayuki Y. Fujimoto\altaffilmark{1}}

\email{suda@astro1.sci.hokudai.ac.jp}

\altaffiltext{1}{Department of Cosmosciences, Graduate School of Science, Hokkaido University, 
Kita 10 Nishi 8, Kita-ku, Sapporo 060-0810, Japan} 
\altaffiltext{2}{Astrophysics Group, EPSAM, Keele University, Keele, Staffordshire ST5 5BG, UK}
\altaffiltext{3}{present address: National Astronomical Observatory of Japan, Osawa 2-21-1, Mitaka, Tokyo 181-8588, Japan}
\altaffiltext{4}{Institute for the Physics and Mathematics of the Universe, University of Tokyo, 5-1-5 Kashiwanoha, Kashiwa 277-8583, Japan}








\begin{abstract}
We investigate the quantitative constraint on the triple-$\alpha$ reaction rate based on stellar evolution theory, motivated by the recent significant revision of the rate proposed by nuclear physics calculations.
Targeted stellar models were computed in order to investigate the impact of that rate in the mass range of $0.8 \leq M / \msun \leq 25$ and in the metallicity range between $Z = 0$ and $Z = 0.02$.
The revised rate has a significant impact on the evolution of low- and intermediate-mass stars, while its influence on the evolution of massive stars ($ M \ga 10 \msun$) is minimal.
We find that employing the revised rate suppresses helium shell flashes on AGB phase for stars in the initial mass range $0.8 \leq M / \msun \leq 6$, which is contradictory to what is observed.
The absence of helium shell flashes is due to the weak temperature dependence of the revised \triplea\ cross section at the temperature involved.
In our models, it is suggested that the temperature dependence of the cross section should have at least $\nu > 10$ at $T = \pow{1 \hyp 1.2}{8}$ K where the cross section is proportional to $T^{\nu}$.
We also derive the helium ignition curve to estimate the maximum cross section to retain the low-mass first red giants.
The semi-analytically derived ignition curves suggest that the reaction rate should be less than $\sim 10^{-29}$ cm$^{6}$ s$^{-1}$ mole$^{-2}$ at $\approx 10^{7.8}$ K, which corresponds to about three orders of magnitude larger than that of the NACRE compilation.
In an effort to compromise with the revised rates, we calculate and analyze models with enhanced CNO cycle reaction rates to increase the maximum luminosity of the first giant branch.
However, it is impossible to reach the typical RGB tip luminosity even if all the reaction rates related to CNO cycles are enhanced by more than ten orders of magnitude.
\end{abstract}





\section{Introduction}

Nuclear reaction rates are fundamental to the modeling of stars.
Among them, helium-burning reactions by the triple-$\alpha$ process play a critical role in evolution of red giants.
The cross section of the triple-$\alpha$ reaction is largely influenced by resonances in the energy states of $\nucm{4}{He}$, $\nucm{8}{Be}$ and $\nucm{12}{C}$.
These resonances were proposed by \citet{Opik1951} and \citet{Salpeter1952};
the resonance level of \nucm{12}{C} was predicted by \citet{Hoyle1953} and \citet{Hoyle1954} to be at $7.68$ MeV, confirmed experimentally with a value of $7.68 \pm 0.03$ MeV \citep{Dunbar1953}.
The contribution by non-resonant reaction was also taken into account by \citet{Nomoto1985} to investigate the impact on helium ignition in accreting white dwarfs and neutron stars.
Their result significantly increases the triple-$\alpha$ reaction rate by more than 20 orders of magnitude at $T \lesssim \pow{3}{7}$ K.
However, this drastic change of nuclear reaction rates do not affect on the main characteristics of stellar evolution because the rate is not so significantly modified within the temperature range for helium ignition in stars ($T \gtrsim \pow{6}{7}$ K).
The only one exception is the behavior of the evolution of low-mass metal-free stars \citep{Fujimoto1990,Suda2007}.
Without CNO elements initially, the hydrogen flash is triggered by CNO cycles as a consequence of the accumulation of carbon in the hydrogen-burning domain through the \triplea.
However, this phenomenon cannot currently be observed since the probability of detecting this event is significantly small from the viewpoint of both timescale and statistics.
Thus the reaction rate by \citet{Nomoto1985} has been adopted by the later compilations of nuclear reaction rates \citep{Caughlan1988,Angulo1999} in place of previous work, which did not include the contribution of non-resonant reactions \citep{Fowler1975}.

Recently, \citet{Ogata2009} (OKK, hereafter) presented results on the cross section of the triple-$\alpha$ reaction using a direct calculation of Schr\"odinger equation.
They also provide a tremendously larger cross section at $10^{7 \hyp 8}$ K compared with the previous compilations for this nuclear reaction rate.
The difference amounts to 26 orders of magnitude around $10^7$ K compared with the NACRE compilation \citep{Angulo1999}, i.e., the reaction rates by \citet{Nomoto1985}, and also with the latest reaction rate \citep{Fynbo2005}.
This result gives rise to a severe inconsistency with the current understandings of the observations.
In particular, low-mass red giants are no longer produced using the revised triple-$\alpha$ reaction rate due to the premature ignition of helium at the stellar center upon arrival at the base of the red giant branch \citep{Dotter2009}.
In addition to this report, \citet{Saruwatari2010}, \citet{Matsuo2011}, and \citet{Peng2010} investigate the impact on accreting C-O white dwarfs and neutron stars.
For the accretion onto white dwarfs, it is suggested that the energy released during off-center helium detonation attained when employing the OKK rate dominates over that of the central carbon ignition that is responsible for Type Ia supernovae.
On the other hand, it is reported that accretion onto neutron stars with the OKK \triplea\ rate results in burst energies from ultracompact X-ray binaries that are too low to account for those observed.
\citet{Matsuo2011} insist that the OKK rate should be reduced by a factor of $10^{2-3}$ for $10^{8} {\rm K}$ to be consistent with the observations of X-ray bursts.
Application of the OKK rate to Cepheid models by \citet{Morel2010} also fails to explain the accumulations of observations reporting the period change, depletion of lithium and reduction of the C/N ratio, although they insist that the OKK rate may reduce the mass discrepancy between stellar evolution models and pulsation theory.

Stimulated by this revision of the \triplea\ rate, new calculations of \triplea\ rates such as a method solving the Faddeev three-body equations have been ongoing (S.~Ishikawa, private communication).
It is also proposed that broad resonances other than the Hoyle state exist, which may also affect the \triplea\ at low temperatures \citep[see, e.g.,][]{Kato2010}.
On the other hand, \citet{Ogata2009} point out that the treatment of non-resonant reactions at low-temperatures adopted by \citet{Nomoto1985} is not appropriate.
Therefore, it is desired to explore the reaction rates both theoretically and experimentally, and hence, it is likely that the cross section for \triplea\ will be modified by future works.
As a calibration test for the modification of the reaction rate, it is useful to rely on stellar evolution models by analyzing the dependence of the theoretical models on the reaction rate, especially below $\pow{3}{8}$ K where the influence is important for the evolution of low- and intermediate-mass stars.

In this paper, we provide quantitative constraints on the \triplea\ cross section based on stellar evolution theory.
Firstly, we investigate the impact of changing the triple-$\alpha$ reaction rate on the evolution of low- and intermediate-mass stars during the horizontal branch and AGB phase, extending our study to the evolutionary characteristics of massive stars.
The possible effects of changing the \triplea\ rate are analyzed by comparing our models with the observed extremely metal-poor stars in the Galaxy using the SAGA database \citep{Suda2008}.
Secondly, we derive the possible range of the \triplea\ rate by estimating the ignition point of helium burning on the first red giant branch (FGB) in a semi-analytic method.
We also explore the possibility of varying the input physics with the OKK \triplea\ rate adopted.
In particular, we discuss the change of the reaction rates of CNO cycles.

The paper is organized as follows.
In the next section, the stellar evolution models used in this work are described.
The impact of changing the \triplea\ rates is also discussed.
\S~\ref{sec:ignition} is devoted to the derivation of helium ignition curve and the comparison with observations.
In \S~\ref{sec:cnocycle}, we look for a compromise with the OKK \triplea\ rate by varying other input physics.
Summary and conclusions follow in \S~\ref{sec:sum}.

\section{Stellar evolution with the OKK rate}\label{sec:evo}
The stellar evolution code and adopted input physics are the same as in \citet{Suda2010} except for conductive opacities for which we adopt the calculation by \citet{Itoh2008}.
For the nuclear reaction rates, we use both the NACRE compilation \citep{Angulo1999} and the OKK rate.
The mixing length parameter $\alpha$ is set at 1.5.
We followed the evolution from zero-age main sequence to the early phase of the AGB.
Initial composition is scaled solar abundance based on \citet{Anders1989} without any $\alpha$-element enhancement.
We also discuss the effect of using the OKK rate on the evolution of massive stars and present the results with the model of 12 and $25 \msun$ with $Z = 0.01$ using the Geneva code \citep{Hirschi2004}.
For these models, $\alpha = 1.6$ is adopted.
For massive star models, calculations are terminated at the beginning of carbon burning phase.
Hereafter we refer to the models computed with NACRE rates as fiducial models.

\begin{figure}
  \plotone{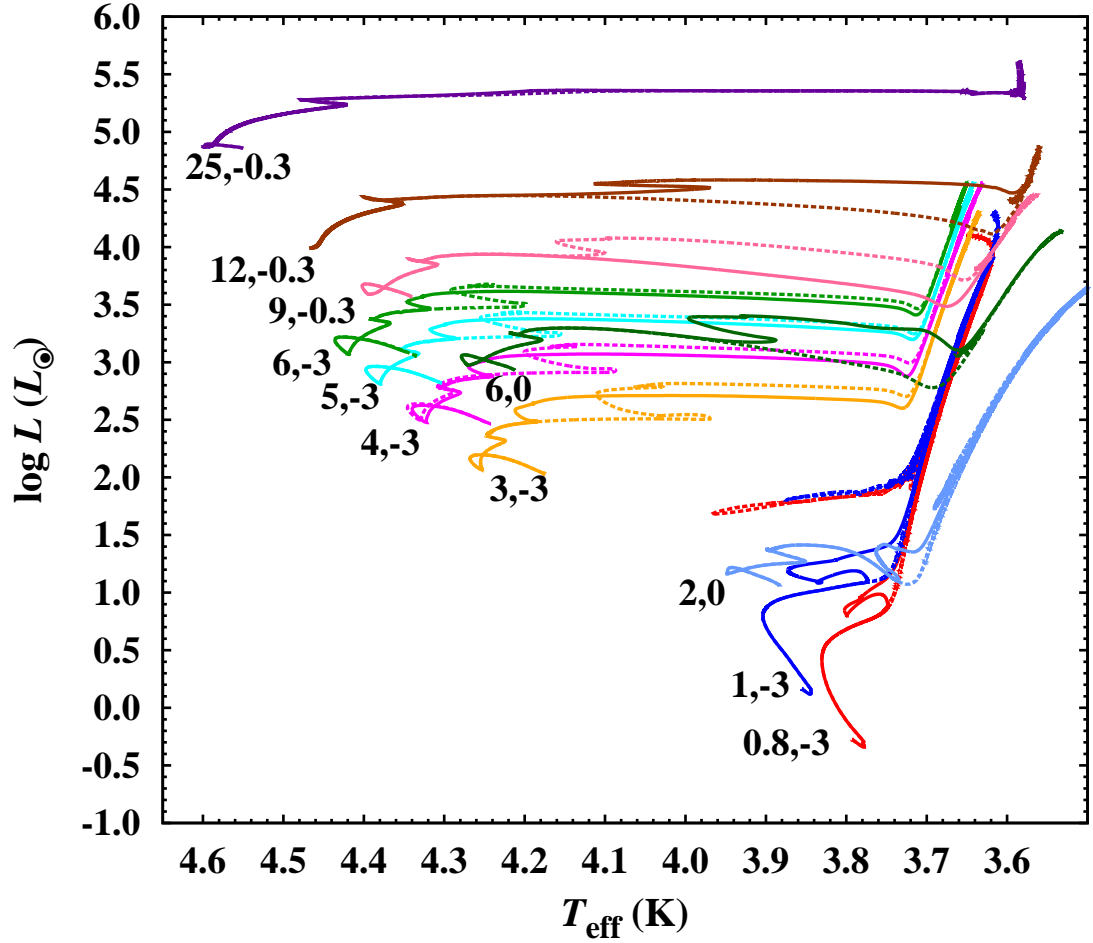}
  \caption{Comparison of selected evolutionary tracks on the H-R diagram using the OKK (solid lines) and NACRE (dashed lines) rate.
  Initial mass (in \msun) and metallicity (in \feoh) of stars are labeled on the next to the main sequence.
  For $12$ and $25 \msun$ with $\feoh = -0.3$, we plot the models computed with the Geneva code.
  }
  \label{fig:hrd}
\end{figure}

Figure~\ref{fig:hrd} compares evolutionary tracks for various sets of mass and metallicities using the NACRE and the OKK rate.
For models of $M \gtrsim 10 \msun$, the impact of adopting the OKK rate is not significant since more massive stars ignite helium at higher temperatures and lower densities.
At the start of core He burning, the nuclear burning rate is larger in the OKK rate than in the NACRE rate for a $12 \msun$ model by a factor of $\sim 10^{3}$, while it is only larger by a factor of 5 to 10 for a $25 \msun$ model.

The helium ignition temperature is $\pow{1.14}{8}$ K (\pow{1.55}{8} K with the NACRE rate) and $\pow{1.52}{8}$ (\pow{1.75}{8}) K for $12 \msun$ and $25 \msun$ respectively, and is weakly dependent on the \triplea\ rates adopted for this mass range.
The OKK rate gives larger luminosity for the $12 \msun$ model, placing it higher in the H-R diagram because helium burning starts in contraction phase, while it occurs on red giant branch for the fiducial model.
The OKK rate also gives larger final \nucm{12}{C} mass fraction at the end of core helium burning.
A higher \nucm{12}{C} mass fraction can change the size of the convective cores during the advanced burning stages and also the final iron core mass.
For $25 \msun$, such differences are negligible.

For low- and intermediate-mass stars, the effect of using the OKK rate is significant even after core helium burning.
As pointed out by \citet{Dotter2009}, the helium ignition occurs in the center when the star is at the base of the FGB on the H-R diagram for an initial mass of $0.8 \msun$, while a helium flash occurs off-center for the initial mass of $M / \msun \lesssim 2$ for the fiducial models, with some dependence on metallicity \citep[see, e.g.][]{Suda2007b,Suda2010}.
In these low-mass stars, the luminosity at the horizontal branch phase is smaller by a factor of $3 \hyp 60$ compared with fiducial models because of small helium core masses ($\sim 0.2 \hyp 0.25 \msun$) at the point of helium ignition.
Because of the low luminosity, the duration of the core helium burning can be longer than the fiducial models.
Since the typical lifetime on horizontal branch is $50 \hyp 100$ Myrs, low-luminosity horizontal branch stars should be detected in the observations of globular clusters if the OKK rate is adopted.
Once the core has exhausted helium in the center, the star ascends the red giant branch for the first time.
This can be interpreted as red giants, but there is a significant gap between the helium ignition point and the ascension of the giant branch, which has never been detected in the observed color-magnitude diagrams.

\begin{figure}
  \plotone{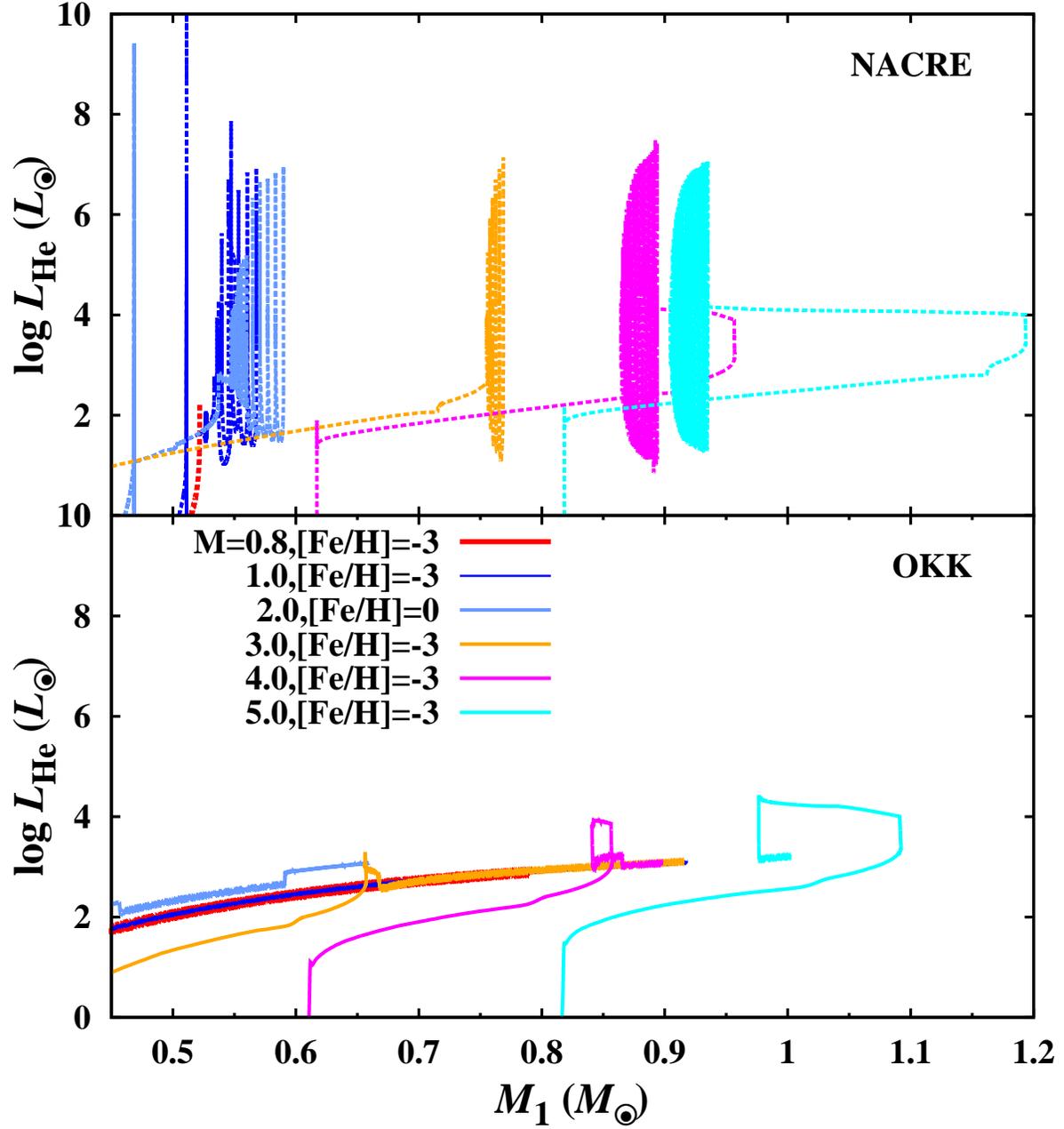}
  \caption{Comparison of helium burning luminosity as a function of helium core mass for the models with the NACRE rate (top panel) and the OKK rate (bottom panel).
  The definition of the lines is the same as in Fig.~\ref{fig:hrd}, but the models undergoing carbon burning ($M \ge 6 \msun$ with the OKK rate) are excluded.
  }
  \label{fig:agb}
\end{figure}

\begin{figure}
  \plotone{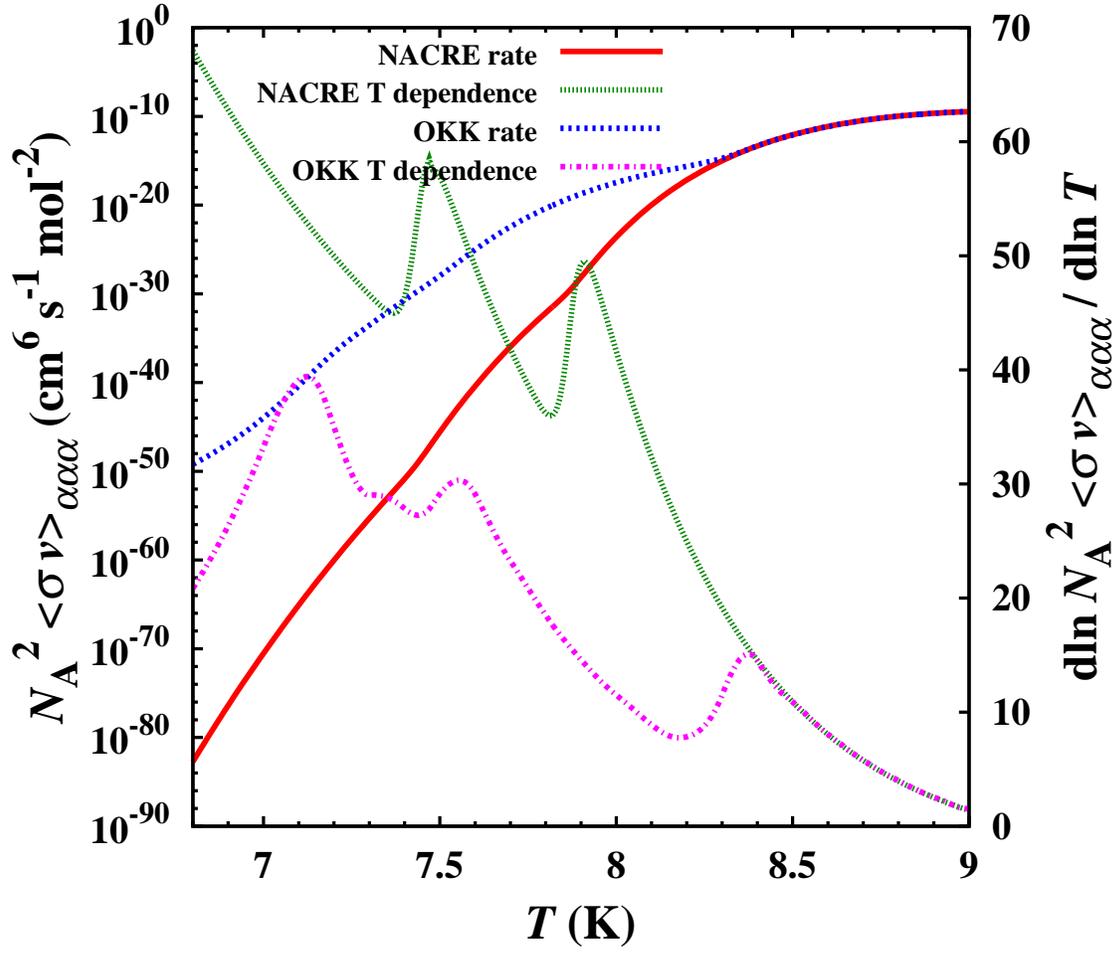}
  \caption{Cross sections for triple-$\alpha$ reactions and their temperature dependences ($d \ln \langle \sigma\,v \rangle_{\alpha \alpha \alpha} / d \ln T$, see right axis) for the NACRE rate and the OKK rate.
  }
  \label{fig:tdep}
\end{figure}

Another remarkable difference is that the models with $M \geq 0.8 \msun$ do not experience thermal pulsations of helium shell burning during the AGB phase.
Figure~\ref{fig:agb} shows the variations in helium burning luminosity as a function of helium core mass during AGB evolution.
Models are taken with the NACRE or OKK rate for the mass range of $0.8 \hyp 5 \msun$.
In the OKK models, helium shell-burning occurs in a radiative zone with a front that advances steadily without thermal pulses.
This leads to a smooth, gradual increase of the carbon-oxygen core mass together with a gradual increase of the helium burning luminosity.
The absence of thermal pulses or shell-flashes is a consequence of the weaker temperature dependence of triple-$\alpha$ cross sections for the OKK rate.
Figure~\ref{fig:tdep} compares the temperature dependence, $\nu$, of \triplea\ rates, defined by $\nu \equiv d \log \langle \sigma v \rangle_{\alpha \alpha \alpha} / d \log T$, as well as the cross section itself.
The value of $\nu$ for the OKK rate is smaller than $\approx 8$ for the temperature range for helium shell burning ($\pow{1 \hyp 1.2}{8}$ K) in the present computations.
The absence of thermal pulses is a serious challenge to observations.
Without thermal pulses, AGB models do not experience any third dredge-up events that are responsible for the enhancement of carbon and \sit-process elements.
Therefore, the mass transfer scenario, which is likely to account for the carbon-enhancement observed in CH and CEMP stars (including CEMP-\sit\ stars), is not viable \citep[see, e.g.][]{Suda2011}.

The operation of the third dredge-up is also supported by the detection of \nucm{99}{Tc} in AGB stars \citep{Merrill1952,Smith1983}.
Since the half-life of \nucm{99}{Tc} is only \pow{2.13}{5} years, this element is thought to be produced by the \sit-process and dredged-up to the surface during each thermal pulse cycle \citep{Cameron1955}.
In addition to avoiding dredge-up events, we also exclude the possibility for efficient mass loss, such as super wind, triggered during the late stage of AGB evolution because of the lack of carbon that can be condensed into dust to drive the wind mass loss \citep{Iben1983b,Bowen1991,Lagadec2008}.
Without the efficient mass loss, AGB stars can not evolve to white dwarfs, and hence explode as supernovae Type~I~$1/2$ when the core mass approaches the Chandrasekhar mass limit \citep{Iben1983b}.

With regards to the subsequent stage of evolution, the models with the OKK rate ignite carbon for smaller initial mass ($M \ge 6 \msun$) compared with the models with the NACRE rate ($M \ge 8 \msun$) \citep[see Fig.1 of][]{Suda2010}.
This is because the OKK models have a larger core mass after the second dredge-up (see Fig.~\ref{fig:agb}) due to the earlier helium ignition during the contraction phase after the helium core exceeds the Sch\"onberg-Chandrasekhar mass limit.
In addition, the carbon-oxygen core also becomes massive enough to ignite carbon in the core before the second dredge-up reduces the helium-rich layer.
In our models of $7$ and $9 \msun$ with $Z=0.01$, the mass of the carbon-oxygen core becomes as large as $1.07$ and $1.27 \msun$ when the OKK rate is adopted, respectively.
These are in contrast with the models with the NACRE rate for which we find $0.89$ and $1.01 \msun$ for $7$ and $9 \msun$ models respectively.
Since the carbon-oxygen core is weakly degenerate for $7 \la M / \msun \la 9$ model with the OKK rate, these models will become so-called super-AGB stars \citep{GarciaBerro1994}.
We also ran models with the MESA code \citep[Samuel Jones, private communication]{Paxton2011} to study the lower mass limit for core-collapse supernovae and obtained similar results concerning the impact of the \triplea\ rate on the boundary mass for carbon burning.
Based on the models we have calculated, we estimate that the OKK rate decreases the minimum mass for core-collapse supernova by up to $\sim 2 \msun$.

Finally, we comment on the comparisons of our models with those computed by \citet{Morel2010}.
Our models of $4 \hyp 6 \msun$ stars are consistent with those of \citet{Morel2010}, although the metallicity range is different.
We also computed $6 \msun$ models with solar composition to confirm the blue loops on the RGB.
We find two blue loops during the central helium burning and during the growth of $\rm{C} + \rm{O}$ core, though the size of the loop is smaller than their models.
For $\feoh = -3$, both blue loops are not observed because the helium burning starts during the contraction of the helium core on the subgiant branch, which prevent from the appearance of the first loop.
The second loop appears when the mass of the $\rm{C} + \rm{O}$ core is smaller than Sch\"onberg-Chandrasekhar mass limit, i.e., $\approx 0.37$ times the mass of the helium core.
For low metallicity stars, the $\rm{C} + \rm{O}$ core mass exceeds this limit on the RGB, which results in the absence of another loop.
If we adopt the OKK rate, our $6 \msun$ model agrees qualitatively well with their $6.6 \msun$ model with the OKK rate.

\section{Effect of the triple-$\alpha$ reaction rates on helium ignition}\label{sec:ignition}
In the previous section, we see significant modifications to the current understanding of stellar evolution by adopting enormously enhanced \triplea\ rates.
Here we want to establish constraints on the \triplea\ rate based on the occurrence of the FGB, which is supported by observations.
We derive the maximum cross section for the \triplea\ by considering its effects on the temperature of helium ignition, which corresponds to the maximum luminosity of the FGB.
In the following, we derive the helium ignition point from stellar evolution theory and models based on \citet{Fujimoto1981,Fujimoto1982b,Fujimoto1982c} and \citet{Suda2010}, respectively, and estimate the allowed range of the \triplea\ rates that is compatible with the observations.

The ignition of helium shell burning is discussed by \citet{Fujimoto1981} and its occurrence is described in eq.~(9).
Here we may simply define the helium ignition point by the requirement that helium ignition occurs in a shell where the nuclear timescale reduces to be as small as the thermal diffusion timescale.
According to eq.(7) in \citet{Fujimoto1981}, the diffusion timescale is in inverse proportion to the rate of diffusion, $\epsilon_{r}$, which is given by
\begin{eqnarray}
\epsilon_{r} &=& \frac{4ac}{3} \frac{T^{4}}{\kappa H_{P}^{2} \rho^{2}} \\
             &=& \frac{4acG}{3} \frac{M_{r}^{2} T^{4}}{\kappa r^{4} P^{2}}
\end{eqnarray}
where $H_{P}$ is the pressure scale height and the other symbols have their usual meanings.
In deriving the condition for helium ignition, we assume that the rate of diffusion is identical to the rate of nuclear energy generation, $\epsilon_{n}$, at the shell potentially producing the largest nuclear energy, i.e., the shell at the highest temperature.
Therefore, we can obtain the ignition cross section given by the replacement of $\epsilon_{n}$ with $\epsilon_{r}$ in the expression for the cross section of the \triplea,
\begin{eqnarray}
N_{\rm A}^{2} \langle \sigma v \rangle_{gs}^{\alpha \alpha \alpha} &=& \frac{1}{N_{\rm A} Y_{4}^{3}} \frac{6}{\rho^{2}} \frac{\epsilon_{r}}{E_{\alpha \alpha \alpha}}, \label{eq:cs}
\end{eqnarray}
where $Y_{4}$ is the number abundance of \nucm{4}{He}, and $E_{\alpha \alpha \alpha}$ the Q value of \triplea, which is set at 7.274 MeV in this study according to \citet{Angulo1999}.

\begin{figure}
  \plotone{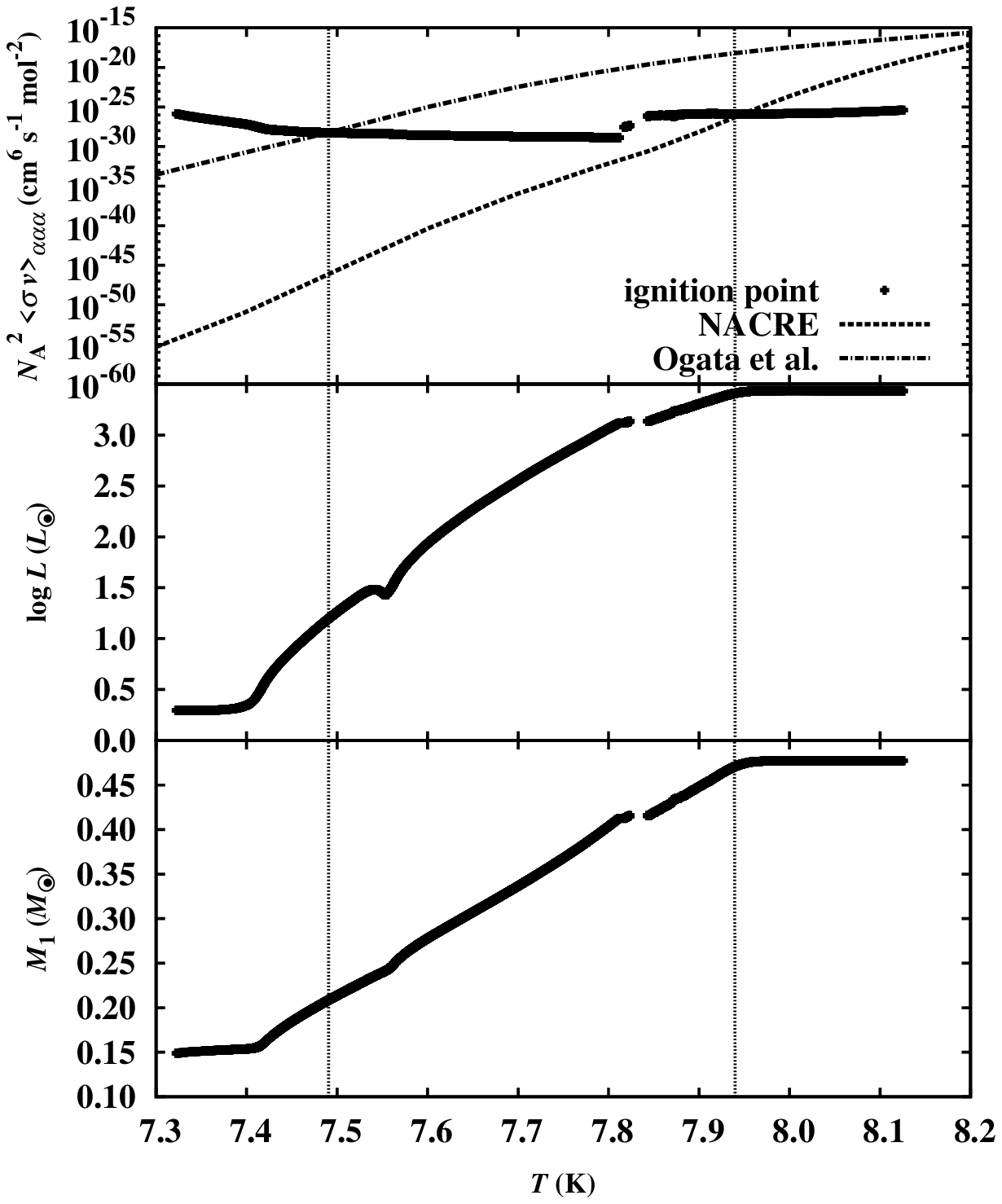}
  \caption{
		Model result of $1 \msun$ with $Z=0.02$.
		Top panel shows the comparison of the expected ignition points computed by eq.~(\ref{eq:cs}) with the cross sections by \triplea\ taken from NACRE (broken line) and \citet[dash-dotted line]{Ogata2009}, as a function of temperature at the shell where the \triplea\ reaction rate is maximum.
		Middle and bottom panel are the luminosity and helium core mass, respectively.
		Dotted vertical lines represent the intersection between the reaction cross sections and ignition curves determined from stellar models.
  }
  \label{fig:ignition}
\end{figure}

The top panel in Figure~\ref{fig:ignition} shows an example of the ignition curve for $1 \msun$ model with $Z = 0.02$ using the standard set of NACRE reaction rates.
Each point denotes the helium ignition point as a function of the maximum temperature in the model from the onset of hydrogen shell burning to the beginning of the core helium flash.
The maximum temperature occurs at the center until the core mass grows larger than $M_{1} = 0.404 \msun$ for $\log T_{\rm max} = 7.80$, as shown by a break in the lines.
After that, it is shifted to the outer shell as the central region is cooled effectively via neutrino losses and develops a temperature inversion.
The corresponding luminosity and helium core mass are shown in middle and bottom panels respectively.
This model experiences a helium shell flash at the mass co-ordinate $M_{r} = 0.230 \msun$ when $M_{1} = 0.477 \msun$. 
The density and temperature in the igniting shell are $10^{5.46}$ g cm$^{-3}$ and $10^{7.94}$ K respectively.
As shown in the figure, the ignition point is well predicted by the intersection of the ignition curve and nuclear reaction cross section provided by the NACRE compilation (dotted curve) as shown by the right vertical line in the figure.
The prediction for the ignition point also holds for the OKK rate (dot-dashed curve).
The derived ignition point for this model suggests that the helium ignition occurs at the temperature of the shell with $T \approx 10^{7.49}$ K if we adopt the OKK rate (see dotted vertical line in the left).
The corresponding luminosity and core mass are $\log (L/\lsun) \approx 1.1$ and $M_{1} \approx 0.2 \msun$ respectively, shown in the middle and bottom panel of Figure~\ref{fig:ignition} and emphasized by the vertical line.
These values are in quite good agreement with the results of the same model computed by \citet{Dotter2009}.

\begin{figure}
  \plotone{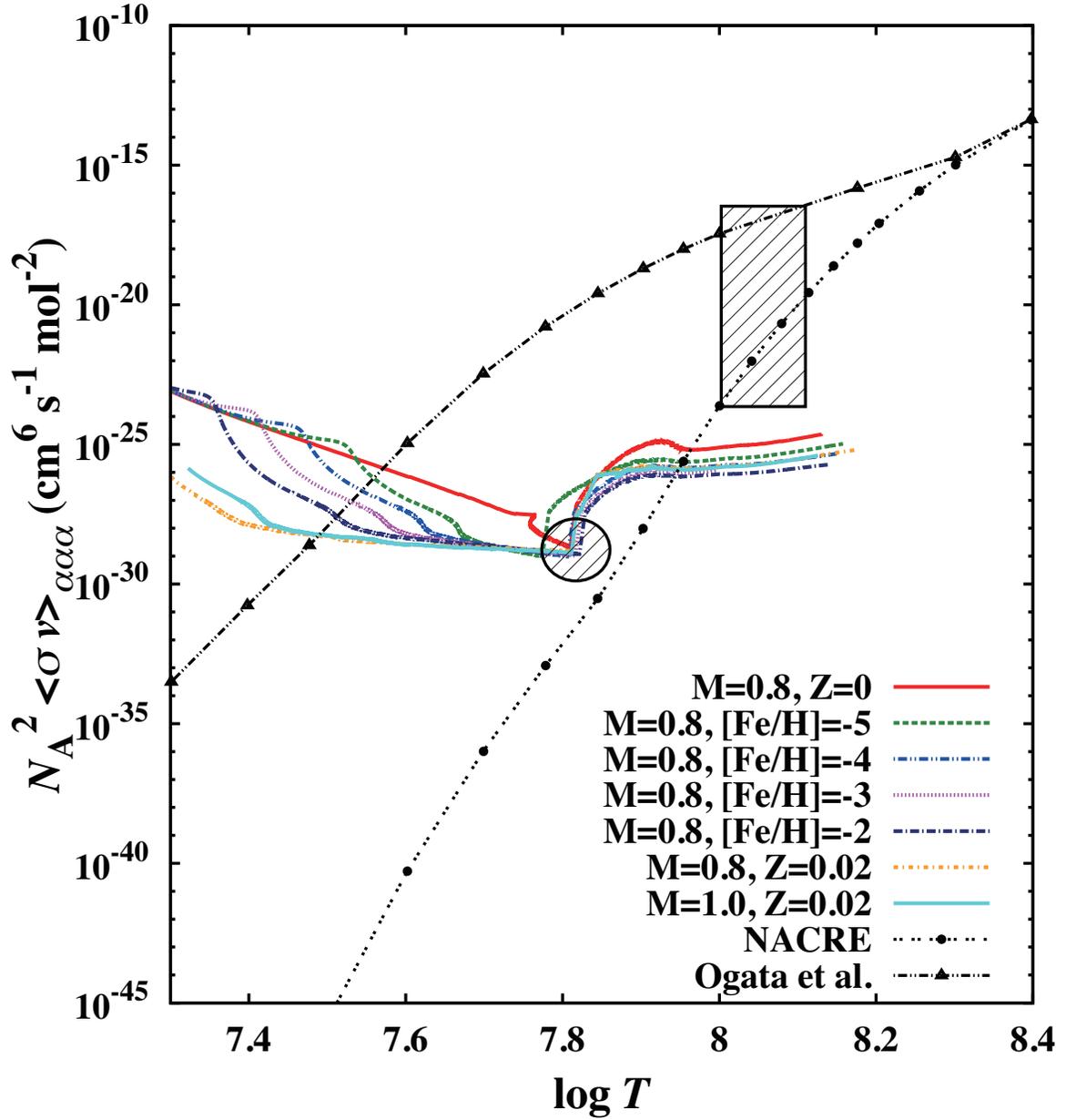}
  \caption{Ignition curves derived by our stellar models compared with the cross section of \triplea\ of NACRE and \citet{Ogata2009}.
  	    Hatched areas denote the constraints on the reaction rate obtained by this study.
		Hatched circle designates the maximum cross section to avoid central helium ignition in low-mass red giants.
		Hatched square shows the temperature range for which the temperature dependence of the cross section should be large enough ($\nu \gtrsim 10$) so that low- and intermediate-mass stars undergo helium shell flashes.
		}
  \label{fig:tria}
\end{figure}

Figure~\ref{fig:tria} compares the ignition lines derived for the models of low-mass
stars with various metallicity and metal-free stars.
The different trajectories at different metallicities in the range $7.4 \lesssim \log T ({\rm K}) \lesssim 7.7$ reflects the different thermal evolution of the core during the subgiant branch where the thermal energy in the hydrogen burning shell is larger than the gravitational energy.
This leads to the variations in the intersection of the cross sections between model predictions and the OKK rate since the mass of helium core differs during the transition from the central hydrogen-burning to the shell hydrogen-burning phase.
Once the star evolves to the RGB, the rate of diffusion in the core is almost independent of total mass and metallicity.
Therefore, the values of the ignition cross section at the center are almost identical to each other at $\log T \gtrsim 7.7$ until the temperature inversion develops and the largest nuclear burning rate takes place in the off-center shells.
A slight difference of the ignition cross section arises after neutrino energy losses become active, which simply results from the difference in the compression rate of the helium core due to hydrogen shell burning.

\begin{figure}
  \plotone{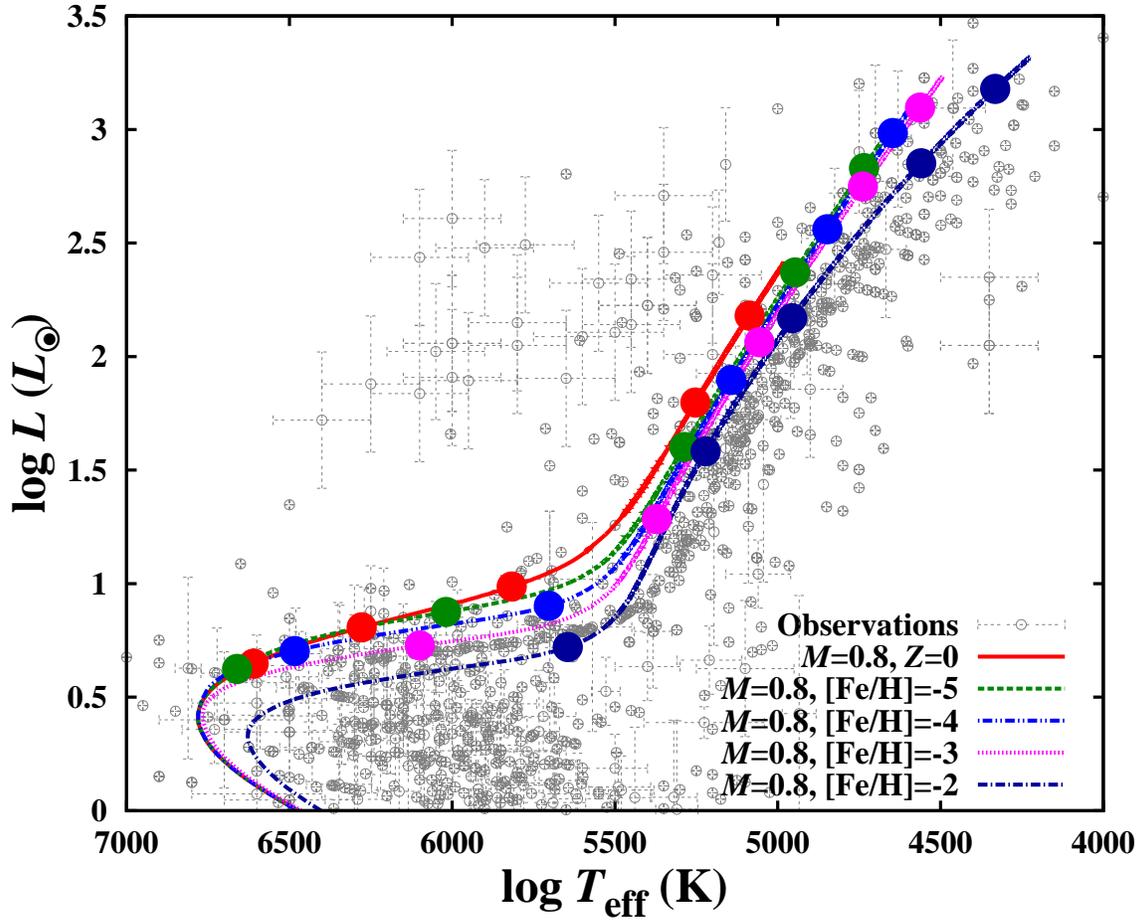}
  \caption{The end point of the FGB for models with the different \triplea\ rate on the H-R diagram.
  Each filled circle on evolutionary tracks shows the ignition point when the ignition temperature is $\log T = 7.5, 7.6, 7.7, 7.8$, and $7.9$ from bottom to top on the same line.
  Observed halo stars are also plotted using the SAGA database \citep{Suda2008}.
  }
  \label{fig:cmd}
\end{figure}

For $0.8 \msun$ stars, the ignition temperature with the OKK rate becomes $\log T ({\rm K}) \approx 7.5 \hyp 7.6$ with weak metallicity dependence.
The corresponding luminosity on the FGB in the H-R diagram can be seen in Figure~\ref{fig:cmd}.
The comparison is made using the SAGA database \citep{Suda2008} that contains the stellar parameters of low-mass halo stars with metallicity mainly in the range $-4 < \feoh < -1$.
The points on each evolutionary track show the helium ignition point at the temperature of $\log T ({\rm K}) = 7.5$ to $7.9$ in $0.1$ dex steps.
In order to be compatible with the observations, the ignition temperature has to be as large as $\log T ({\rm K}) \simeq 7.9$.
Accordingly, we find that the upper limit for the \triplea\ cross section will be at most 2 to 3 orders of magnitude larger than the current rates as given by the NACRE compilation in order to remain coherent with obvsevations, unless there are any other changes to the input physics.
In other words, the shape of the cross section can be arbitrary as a function of temperature as long as it intersects the ignition curve at the sufficiently high temperature to delay the helium ignition.
More specifically, the cross section of the \triplea\ should be less than $\sim 10^{-29}$ cm$^{6}$ s$^{-1}$ mole$^{-2}$ at $\log T \approx 7.8$.
Another condition is that the cross section must have large temperature dependence of $\nu \gtrsim 10$ ($\langle \sigma v \rangle \propto T^{\nu}$) at $8.0 \lesssim \log T \lesssim 8.1$ at least to trigger the shell helium flashes as discussed in the previous section.
These constraints are emphasized by hatched areas in Fig.~\ref{fig:tria}.

\section{Can other uncertainties compensate for the change introduced by the OKK rate?}\label{sec:cnocycle}

As discussed in the previous sections, astrophysical constraints clearly rule out the OKK rate.
In this section, we try to work out if the effects of the very large OKK reaction rate can be balanced by changing other input physics and reaction rates.
In particular, we focus on the reaction rates of the CNO cycle to preserve the FGB with variations in the \triplea\ rate, such as the OKK rate.
Other input physics such as opacities, equation of state, and neutrino loss rates are not worth considering because they will not contribute to dramatic changes of the evolution during the FGB.
At least, there is no supporting evidence that stellar physics should be modified by more than an order of magnitude in the viewpoint of neither theory nor observations.
For nuclear reaction rates, there are non-negligible uncertainties in some reactions, although the reactions related to hydrogen burning are rather well determined and the current rates used in stellar models are compatible with observations.

A possible solution to retain the FGB is the idea that enhancing the CNO cycle reaction rates accelerates evolution during the ascent of the RGB before igniting helium in the core.
Since red giants are supported by hydrogen shell burning mainly by CNO cycles, their enhanced reaction rates be able to increase the luminosity of the star rapidly.
In order to estimate how much enhancement is required to accelerate the FGB evolution up to $\log L / \lsun \sim 3$, we compute models with enhanced reaction rates for the CNO cycles.
The maximum luminosity on the FGB obtained by the model computations is plotted by filled circles in Fig~\ref{fig:rcno}.
One should note that all the reaction rates related to CNO cycles have to be multiplied by the same factor in order to increase the maximum luminosity.
Otherwise the bottle neck reactions prohibit the rapid growth of helium core and terminate the ascent on the FGB.
In our models, the maximum luminosity saturates for the enhancement factors of $10^{10}$ or larger and cannot reach $\gtrsim 10^{3} \lsun$.
Therefore, there is no solution to attain $\log L / \lsun = 3$ as long as the OKK rate is used, even if we adopt enhanced CNO cycle reaction rates much larger than realistic values.
For other metallicities, we estimate the maximum FGB luminosity using a semi-analytic method shown by lines in Fig.~\ref{fig:rcno}, which is discussed in the Appendix.

Figure~\ref{fig:rhrd} shows the computations of $0.8 \msun$ models with $\feoh = -3$ by adopting the OKK rate with enhanced CNO cycle reaction rates.
In addition, our models of low-mass stars with enhanced CNO reaction rates display quite different behaviors on the H-R diagram;
the turn-off point at the main sequence moves toward the right because of the dominance of CNO cycle at the beginning of the main sequence.

\section{Summary and Conclusion}\label{sec:sum}

We investigated the impact of changing the reaction rate of the \triplea\ on stellar structure and evolution.
Models with the reaction rate provided by \citet{Ogata2009} (the OKK rate) are computed for stars in the mass range from $0.8$ to $25 \msun$.
It is shown that the OKK rate brings about several serious problems in astrophysical contexts including the disappearance of the first red giant branch (FGB) as already pointed out by \citet{Dotter2009}.
Here we have found another significant problem that low-mass stars do not experience helium shell flashes due to the weak temperature dependence of the reaction rate.
The absence of helium shell flashes results in the absence of the change of surface chemical composition through the third dredge-ups.
This would suppress important channels to the formation of carbon-enhanced stars since, for a majority of carbon stars, their abundances are best explained by AGB evolution or by binary mass transfer from AGB stars.
In addition, it would also prevent the formation and transportation mechanisms of \sit-process elements in low- and intermediate-mass stars.
As a consequence, AGB stars would no longer be the progenitors of carbon stars and \sit-process element enhanced stars.
In particular for metal-poor halo stars, the large fraction of observed carbon-enhanced stars with \sit-process element enhancement could not be explained by mass transfer from the former AGB stars.
Observationally, the detection of Technetium in the surface of AGB stars is the supporting evidence of the dredge-up of \sit-process elements by the deepening of the convective envelope.
On the other hand, for massive stars with $M \gtrsim 10 \msun$, the effect on the evolution is small because the helium ignition temperature is so high that the difference in the reaction rate between the NACRE \citep{Angulo1999} and the OKK rate is not significant.

We also quantitatively estimated the allowed range of the triple-$\alpha$ cross section under the currently adopted input physics in stellar evolution code.
In our analysis of the helium ignition point of the star, it is suggested that the \triplea\ rate should not be larger than $\sim 10^{-29}$ cm$^{6}$ s$^{-1}$ mole$^{-2}$ at $10^{7.8}$ K in order for stars to still have a red giant phase.
In addition, it is shown that the occurrence of helium shell flashes requires the temperature dependence of the \triplea\ cross section with at least $\nu \gtrsim 10$ at $T \sim \pow{1 \hyp 1.2}{8}$ K where the cross section is proportional to $T^{\nu}$.
For other temperature ranges, the shape of the reaction rate can be arbitrary as a function of temperature only if it intersects the ignition cross section given in eq.(\ref{eq:cs}) at $\log T \gtrsim 7.9$ where helium ignition occurs at the FGB luminosity of $L \gtrsim 10^{3} \lsun$.

A way to compensate for the changes due to the OKK rate was sought by pursuing the increase of the CNO cycle reaction rates, but found to be impossible.
We looked for a solution to increase the FGB tip luminosity by increasing the CNO cycle reaction rates to accelerate the growth of helium core.
Even if we adopt the enhancement factor of $> 10^{10}$ for all the reaction rates related to CNO cycles, the maximum luminosity of the FGB cannot reach $10^{3} \lsun$.

The method to derive the \triplea\ rate proposed by \citet{Nomoto1985} is severely criticized by \citet{Ogata2009}, while the result obtained by \citet{Ogata2009} is still far from being consistent with observations.
Since the \triplea\ is in the wide range of applications such as accretion events on compact stars like novae, X-ray bursts and the determination of cosmological distances through the Cepheid variables, more investigations for this important reaction are desired.



\acknowledgments

The authors are grateful to anonymous referee for useful suggestions to improve the manuscript.
They are deeply indebted to Samuel Jones for proof reading the paper and providing the data computed with the MESA code.
They are also grateful to Kiyoshi Kat{\= o} for helpful comments on cross section for \triplea\ rate.
This work has been supported by Grant-in-Aid for Scientific Research (18104003), from Japan Society of the Promotion of Science.
T. S. has been supported by a Marie Curie Incoming International Fellowship of the European Community FP7 Program under contract number PIIF-GA-2008-221145.
R. H. acknowledges support from the World Premier International Research Center Initiative (WPI Initiative), MEXT, Japan.






\appendix

\section{Semi-analytic method to derive the maximum luminosity of the FGB}

\begin{figure}
  \plotone{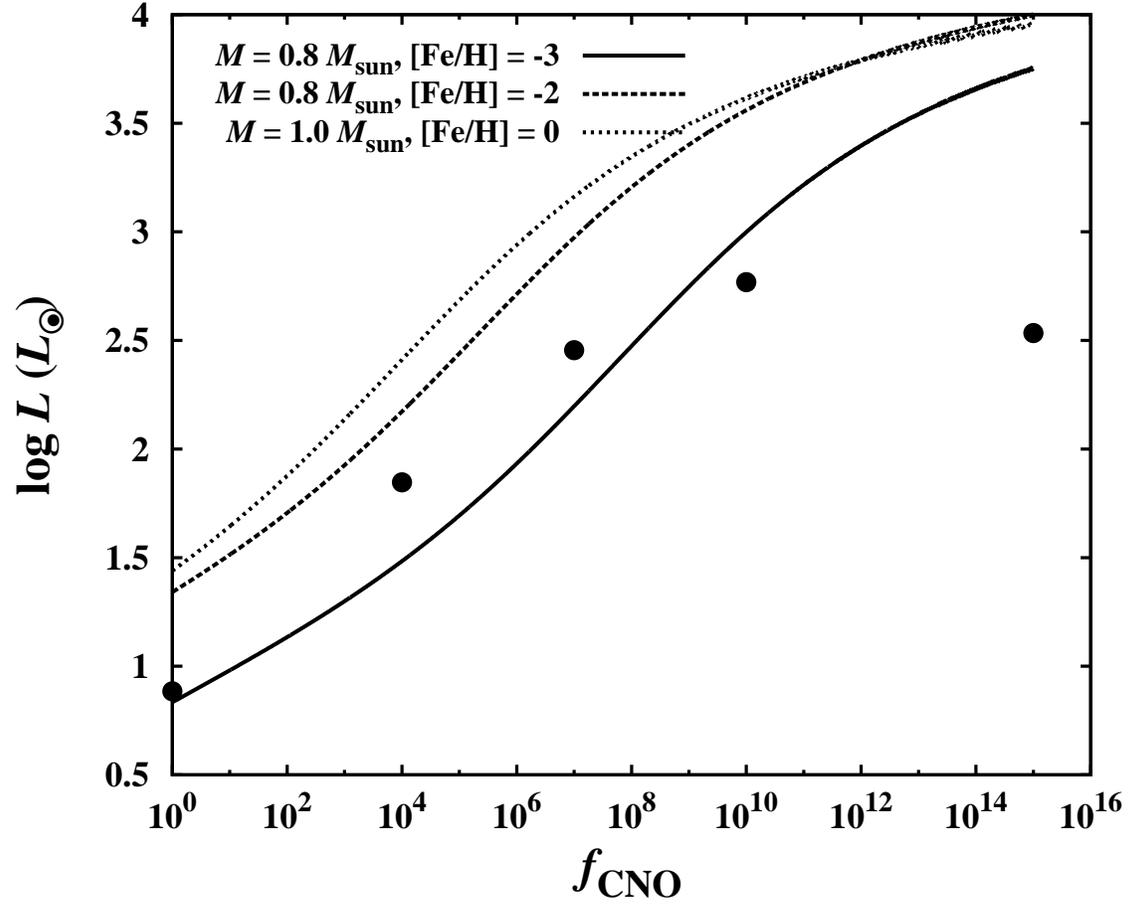}
  \caption{The maximum luminosity in the FGB obtained by multiplying the cross sections of CNO cycles by the factor $f_{\rm CNO}$.
  Each circle denotes the luminosity at the tip of the FGB obtained by the computation of a model of $0.8 \msun$ with $\feoh = -3$ using the OKK rate with enhanced CNO cycle reaction rates.
  Lines show the predictions by our semi-analytic model (see the Appendix).
  }
  \label{fig:rcno}
\end{figure}

As a semi-analytic test to quantify the maximum luminosity at the tip of the FGB, we derive the relation between CNO cycle reaction rates and the maximum luminosity.
Using the expression for nuclear energy generation rate by $\epsilon_{n} \propto \rho^{\eta} T^{\nu}$, the luminosity is given by the following thin-shell approximation \citep{Hayashi1962,Sugimoto1978} using the homologous invariant, $U$ and $V$,  defined by $U = 4 \pi r^{3} \rho / M_{r}$ and $V = G M_{r} \rho / (r P)$, respectively \citep{Hayashi1962}.
\begin{eqnarray}
L_{N} &=& \int_{r_{1}}^{R} 4 \pi r^{2} \rho \epsilon_{N,1} dr \\
&=& \frac{M_{1}}{\frac{3}{4} \left( \eta + 1 \right) + \frac{1}{4} \nu - \frac{3}{V_{1}}} \frac{U_{1}}{V_{1}} \epsilon_{N,1}
\label{eq:lnuc}
\end{eqnarray}
where subscript $1$ denotes the hydrogen-burning shell and $\epsilon_{N}$ the nuclear energy generation rate by CNO cycles.
The total stellar luminosity is approximated by the quantity at the hydrogen burning shell, which is written as
\begin{eqnarray}
L = \frac{16 \pi c G M_{1}}{\kappa_{1}} (1 - \beta_{1}) \frac{d \ln T}{d \ln P}
\label{eq:lstar}
\end{eqnarray}
where $\beta_1$ is the ratio of gas pressure to total pressure at the edge of the helium core.
By equating eq.~(\ref{eq:lnuc}) and (\ref{eq:lstar}), together with the assumption that $V_{1} = 4$ and the equation of state, we obtain a set for solutions of density, temperature, and pressure for a given mass and radius of the helium core, and chemical composition.
The assumption that $V_{1} = 4$ at the hydrogen burning shell is based on the theory of shell burning \citep{Hayashi1962,Sugimoto1978}.
We also assumed the polytropic index of $N = 3$ and the surface chemical composition as the input abundances in solving the equation of state.
The mass, $M_{1}$, and radius, $r_{1}$ of helium core at the ignition of helium burning are taken from the model derived for the OKK reaction rates.
Then, the maximum luminosity, $L_{\rm FGB, tip}^{\rm OKK}$, reached in the FGB with the OKK reaction rate is obtained as a function of enhanced factor, $f_{\rm CNO}$, of CNO cycle nuclear reaction rates by the above procedure, which is shown in Figure~\ref{fig:rcno}.
For a given mass and radius of the helium core, the increase in the luminosity stems solely from the contribution, $1 - \beta_{1}$, of the radiation pressure to the total pressure, as seen from eq.~(\ref{eq:lstar}).
The increase in the dominance of the radiation pressure is realized through the decrease of the density in the burning shell because of the strong temperature dependence of the nuclear burning rate.
Accordingly, the dependence of the maximum luminosity on the enhancement factor of CNO cycle is weak, as shown in Figure~\ref{fig:rcno}.

The prediction obtained by the semi-analytic model shows a good agreement with model computations for the enhancement factor up to $f_{\rm CNO} \simeq 10^{7}$ with the $\log L_{\rm FGB, tip}^{\rm OKK} \lesssim 2.5$.
For still larger enhancement factor, however, the numerical models no longer give an increase in the maximum luminosity on the FGB, differently from the analytical predictions.
The decrease of maximum luminosity in numerical models arises from the expansion of the helium core caused by the decrease in the density in the hydrogen burning shell with a finite temperature, while the value of $r_{1}$ is fixed in the above analytical estimate.




\bibliographystyle{apj}
\bibliography{apj-jour,reference,reference_tria,reference_agb}

\clearpage



\begin{figure}
  \plotone{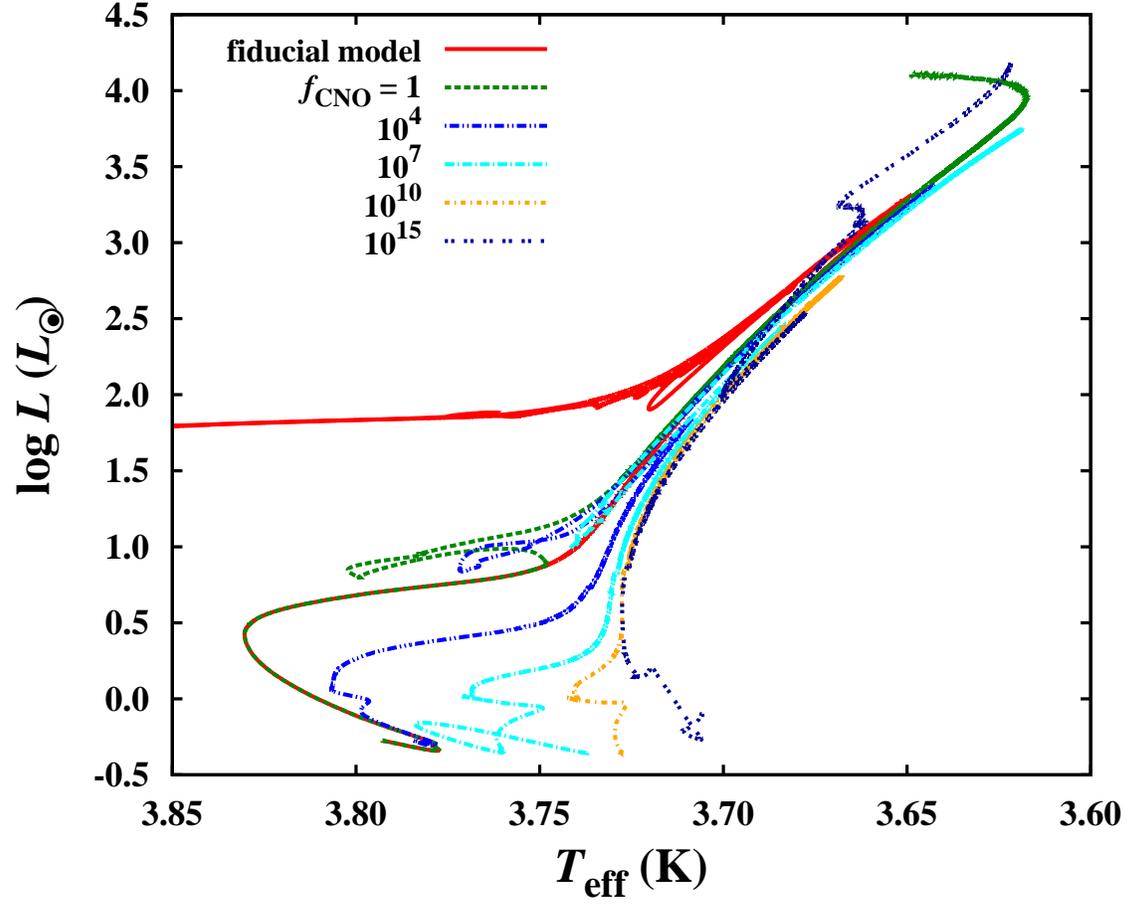}
  \caption{Evolution of $0.8 \msun$ stars and $\feoh = -3$ with the OKK rate and enhanced CNO cycle reaction rates.
  The fiducial model and the models with OKK rate and without the enhancement of CNO cycle reaction rates are plotted for comparison($f_{\rm CNO} = 1$).
  The evolution is followed until the helium core grows nearly to the total stellar mass for the models with enhanced CNO cycle reaction rates.
  }
  \label{fig:rhrd}
\end{figure}

\clearpage

\end{document}